%% file: main.tex
\documentclass[10pt,conference]{IEEEtran}
\makeatletter
\def\ps@headings{%
\def\@oddhead{\mbox{}\scriptsize\rightmark \hfil \thepage}%
\def\@evenhead{\scriptsize\thepage \hfil \leftmark\mbox{}}%
\def\@oddfoot{}%
\def\@evenfoot{}}
\makeatother
\IEEEoverridecommandlockouts
\usepackage{cite}
\usepackage{placeins}
\usepackage{amsmath,amssymb,amsfonts}
\usepackage{algorithmic}
\usepackage{graphicx}
\usepackage{textcomp}
\usepackage{comment}
\usepackage[T2A,T1]{fontenc}
\usepackage{longtable}
\usepackage{lscape}
\usepackage[style=base]{caption}
\usepackage{subcaption}
\usepackage{multirow}

\usepackage[breaklinks]{hyperref} 
\usepackage{url}
\usepackage{breakurl}

    \expandafter\def\expandafter\UrlBreaks\expandafter{\UrlBreaks%  save the current one
    \do\a\do\b\do\c\do\d\do\e\do\f\do\g\do\h\do\i\do\j%
    \do\k\do\l\do\m\do\n\do\o\do\p\do\q\do\r\do\s\do\t%
    \do\u\do\v\do\w\do\x\do\y\do\z\do\A\do\B\do\C\do\D%
    \do\E\do\F\do\G\do\H\do\I\do\J\do\K\do\L\do\M\do\N%
    \do\O\do\P\do\Q\do\R\do\S\do\T\do\U\do\V\do\W\do\X%
    \do\Y\do\Z\do\/\do-}

\usepackage{graphicx}
\usepackage[table,xcdraw]{xcolor}
\usepackage[normalem]{ulem}

\def\BibTeX{{\rm B\kern-.05em{\sc i\kern-.025em b}\kern-.08em
    T\kern-.1667em\lower.7ex\hbox{E}\kern-.125emX}}

\definecolor{darkgreen}{rgb}{0.29, 0.33, 0.13}
\definecolor{std_red}{rgb}{1, 0, 0}
\definecolor{std_green}{rgb}{0, 1, 0}

\newcommand{\system}{Blockly2Hooks}
\begin{document}
\bstctlcite{IEEEexample:BSTcontrol}

%\title{\system: A User-Centric Approach towards Fast and Secure Smart Contract Development 
\title{\system: Smart Contracts for Everyone with the XRP Ledger and Google Blockly
%\thanks{Acknowledgment: we thankfully acknowledge the support from the RIPPLE University Blockchain Research Initiative for our research}
}

%% OLD AUTHORS BLOCK %%%
\begin{comment}
\author{%
\IEEEauthorblockN{Lucian Trestioreanu}
\IEEEauthorblockA{\textit{Interdisciplinary Centre for}\\\textit{Security, Reliability and Trust}\\\textit{University of Luxembourg}\\
Luxembourg, Luxembourg \\
lucian.trestioreanu@uni.lu}
\and \IEEEauthorblockN{Wazen Shbair}
\IEEEauthorblockA{\textit{Luxembourg Company} \\
\textit{Luxembourg}\\ Luxembourg \\ wazen.shbair@gmail.com}
\and \IEEEauthorblockN{Flaviene Scheidt}
\IEEEauthorblockA{\textit{Interdisciplinary Centre for}\\\textit{Security, Reliability and Trust}\\\textit{University of Luxembourg}\\
Luxembourg, Luxembourg \\
flaviene.scheidt@uni.lu}
%\and \IEEEauthorblockN{Aanchal Malhotra}
%\IEEEauthorblockA{\textit{Xpring} \\
%\textit{Northeastern University}\\ Boston, USA \\ %amalhotra@ripple.com}
\and  \IEEEauthorblockN{Radu State}
\IEEEauthorblockA{\textit{Interdisciplinary Centre for}\\\textit{Security, Reliability and Trust}\\\textit{University of Luxembourg}\\
Luxembourg, Luxembourg \\
radu.state@uni.lu}
}
\end{comment}
%%% END OLD AUTHORS BLOCK %%%

%%%%%%%%%COMMENT AUTHORS BLOCK%%%%%%%%%%%%%%%%%%%%%%%%
%\begin{comment}
\author{
\IEEEauthorblockN{
Lucian A. Trestioreanu\IEEEauthorrefmark{1},
Wazen M. Shbair\IEEEauthorrefmark{1},
Flaviene Scheidt de Cristo\IEEEauthorrefmark{1},
%Aanchal Malhotra\IEEEauthorrefmark{2}
and
Radu State\IEEEauthorrefmark{1}
}
 \IEEEauthorblockA{\IEEEauthorrefmark{1} University of Luxembourg, SnT, 29, Avenue J.F Kennedy, L-1855 Luxembourg\\
 Email:\{lucian.trestioreanu, wazen.shbair, flaviene.scheidt, radu.state\}@uni.lu\\
 }
%  \IEEEauthorblockA{\IEEEauthorrefmark{2} Xpring, %Northeastern University, Boston, USA\\
% Email:\{amalhotra\}@ripple.com\\
% }
 }
%\end{comment}
%%%%%%%%% END COMMENT AUTHORS BLOCK%%%%%%%%%%%%%%%%%%%%%%%%

%\author{Anonymous Submission}
%\{lucian.trestioreanu\}@uni.lu}
\maketitle

\input{abstract}
%thttps://www.overleaf.com/project/5e419f39864fb200015c5139

\input{introduction}
%\section{background and related work}
%\label{relatedwork}
\input{background}

\input{design}

\input{related_work}

\input{conclusions}

\section*{Acknowledgment}
%\begin{comment}
%%% GOOD ACK%%%%
We thankfully acknowledge the support of the %XRPL Grants program and the 
RIPPLE University Blockchain Research Initiative (UBRI) for our research.  
%%%%%%%%%%%%%%%%%%%%%%%%%%%
%\end{comment}

%
%%This work was funded in part by the
%This work is supported by the Luxembourg National Research Fund through grant PRIDE15/10621687/SPsquared. In addition, 

%The preferred spelling of the word ``acknowledgment'' in America is without an ``e'' after the ``g''. Avoid the stilted expression ``one of us (R. B. G.) thanks $\ldots$''. Instead, try ``R. B. G. thanks$\ldots$''. Put sponsor acknowledgments in the unnumbered footnote on the first page.
%\newcommand{\cyrshch}[0]{щ}
%\inputencoding{cp1251}
\bibliographystyle{style/IEEEtran}
\bibliography{bib/ebpf}
%\inputencoding{utf8}

\end{document}

%% file: abstract.tex
% !TEX root = main.tex

\begin{abstract}

Recent technologies such as inter-ledger payments, non-fungible tokens, and smart contracts are all fruited from the ongoing development of Distributed Ledger Technologies. The foreseen trend is that they will play an increasingly visible role in daily life, which will have to be backed by appropriate operational resources. For example, due to increasing demand, smart contracts could soon face a shortage of knowledgeable users and tools to handle them in practice. Widespread smart contract adoption is currently limited by security, usability and costs aspects.  Because of a steep learning curve, the handling of smart contracts is currently performed by specialised developers mainly, and most of the research effort is focusing on smart contract security, while other aspects like usability being somewhat neglected. Specific tools would lower the entry barrier, enabling interested non-experts to create smart contracts.%For people without an advanced technical background, smart contracts are currently a kind of mystery. 

In this paper we designed, developed and tested \system, a solution towards filling this gap even in challenging scenarios such as when the smart contracts are written in an advanced language like C. With the XRP Ledger as a concrete working case, \system~helps interested non-experts from the community to learn smart contracts easily and adopt the technology, through leveraging well-proven teaching methodologies like Visual Programming Languages, and more specifically, the Blockly Visual Programming library from Google. The platform was developed and tested and the results are promising to make learning smart contract development smoother.

\end{abstract} 

%% file: introduction.tex
% !TEX root = main.tex

\section{Introduction}
\label{sec:intro}

The recent advances and the growing adoption of the \textit{Distributed Ledger Technology (DLT)} show that DLT is here to stay, with the technology becoming each day more, a part of the daily life. Inter-ledger payments~\cite{spon} made their way into finances~\cite{ilp-foundation}; non-fungible tokens (NFT)~\cite{ante2022non, nadini2021mapping,ante2022eth} are already used in industries like gaming, logistics and more; smart contracts~\cite{kolvart2016smart,zou2019smart} are being successfully deployed to tackle finance, healthcare, gaming, insurance or legal use cases~\cite{nakamoto-szabo-sc}. The term \textit{"smart contracts"} was coined in 1994 by Nick Szabo who defined them as software programs replicating real-life contracts and executing their terms while minimizing exceptions and the need for trusted intermediaries, e.g. notaries~\cite{NSzabo-smart-contracts,DasContract}. Smart contracts~\cite{Szabo1997FormalizingAS} are small pieces of executable code, oftentimes written in advanced programming languages. They reside and execute on DLTs like Ethereum (ETH)~\cite{eth-whitepaper-old,eth-whitepaper-new} or the XRP Ledger (XRPL)~\cite{rippled-exe,xrp-consensus-formal-approach}. Because they exist and run on DLT, they are classified as Decentralized Applications (dApps). Ideally, smart contracts should be autonomous, immutable and public; their execution should be transparent and non-reversible. On ETH, the smart contract's logic is encoded with Solidity, compiled to bytecode, then executed on the Ethereum Virtual Machine~\cite{ETH-Wood-2017}. Solidity is a high level, object-oriented programming language designed specifically to encode smart contracts on Ethereum\footnote{https://docs.soliditylang.org/en/v0.8.18/, valid in February 2023}. Initially inspired from JavaScript, it was enriched with elements from C++ and Python. On XRPL, smart contracts are written in C, compiled to \textit{wasm}\footnote{https://webassembly.org/, valid in February 2023}, then deployed and executed on XRPL. However, the learning curve for smart contract programming is not smooth. Currently, the smart contracts are mostly written, debugged and deployed by specialised developers with advanced technical background. Besides advanced knowledge in the specialized field, another important reason is that smart contracts are written in programming languages like C that can make the process overwhelmingly difficult for non-expert users. Also, the design of the C language is identified as impacting the bug rates, program security and complexity~\cite{Ray-C-complex}. 

Smart contracts should be made friendlier and more accessible to the larger community including teenagers, non-technical people, knowledge seekers and more: in a scenario when smart contracts are deployed on a large scale in industry, it will be desirable that users and stakeholders can directly program the conditions of the contract themselves. Bringing smart contracts closer to the business will unlock their full potential to create value for the society. It is expected that if adopted on a large scale, the financial impact of smart contracts would be substantial. However, their widespread adoption is currently hampered by costs, usability and security~\cite{purnell2022towards}. The specialised programmers needed for development increase the deployment costs. To enable widespread adoption, costs can be decreased by increasing smart contract usability, currently impacted by the advanced programming languages used in development that translate into an entry barrier for non-expert audience. Moreover, when smart contracts are developed by pure programmers, situations can arise when the programmer does not fully understand the needs of the stakeholder, that should be embedded into the contract~\cite{industry-construction-users-themselves}. The stakeholder needs to test the implementation and provide feedback to the programmer making the process iterative and prone to implementation flaws, which increases the costs even further. Currently, there is a general trend towards no-code or low-code tools~\cite{weblow,bubble} for software development to benefit ubiquitous, daily-business use cases~\cite{no-lo-code}. The blockchains whose smart contracts will be more usable and cost-friendly will see their adoption and user-base grow.

Of the possible solutions, we focused on the well-proven \textit{Visual Programming Languages (VPL)} methodology, successfully used in schools to teach programming. VPLs, or \textit{block coding}, have been defined as programming languages that enable users to build programs through interacting with, and expressing the program through, graphical language elements rather than textual~\cite{VPL-roberta-another-one}. Oftentimes they are based on boxes, lines and arrows which interconnect the boxes and represent their relations and interactions~\cite{boxes-arrows1,diagram-VPL}. VPLs aim to make programming more accessible to novices through \textit{syntax, semantics and pragmatics}~\cite{VPL-wiki}. \textit{Syntax} offers icons, blocks, arrows, forms and diagrams to enable beginners to \textit{easily} build \textit{well-formed} programs. \textit{Semantics} are helper methods aiming to convey users the meaning and correct usage of the graphical primitives of the VPL. \textit{Pragmatics} are possibilities offered to the user to test and understand the behavior of pieces of program they create, in different specific situations. VPLs can be used for a variety of domains like multimedia, education, gaming or automation.
\vspace{-4mm}
\begin{figure}[htbp]
    \centering
    \begin{subfigure}[t]{1\columnwidth}
        \centering
        \includegraphics[height=2.2in]{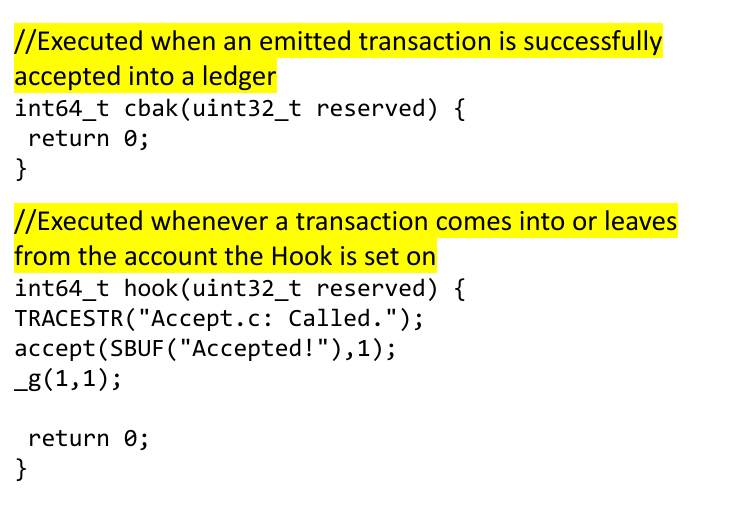}
        \vspace{-5mm}
        \caption{\textit{XRPL} smart contracts, as written by programmers.}
        \label{fig:what-we-have}
    \end{subfigure}
    \newline
    \begin{subfigure}[t]{1\columnwidth}
        \centering
        \includegraphics[height=1.55in]{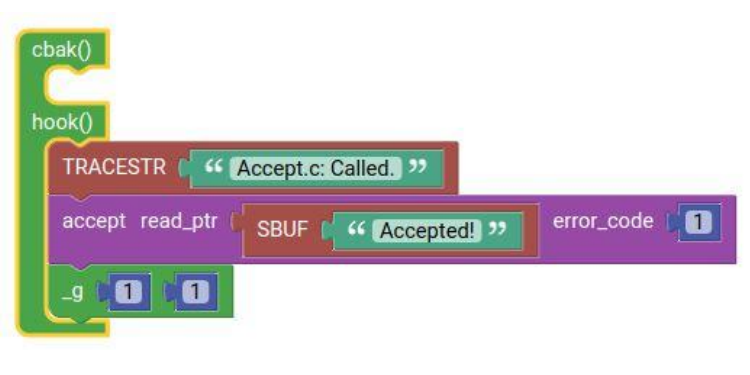}
        \caption{XRPL smart contracts, as developed by non-technical users.}
        \label{fig:what-we-want}
    \end{subfigure}
    %\newline
    %\begin{subfigure}[t]{\columnwidth}
    %    \centering
    %    \includegraphics[width=\linewidth]{figures/Announce-pull/star/new/p-JkCp_time.png}
    %    \caption{Time series: validation interarrival time}
    %    \label{fig:Ann-Pull-vals-inter-star}
    %\end{subfigure}
    \caption{The design goal of~\system.}
    \label{fig:goal}
\end{figure}

Visual editors for smart contracts can democratize the building, deployment and management of smart contracts and improve their reliability and security~\cite{marks}, because, like any other computer program, sometimes smart contracts can have bugs that can lead to potentially significant financial losses~\cite{DAO-hack}. What we want to achieve is illustrated in Figure~\ref{fig:goal}: the complex task to write smart contracts in C (Figure~\ref{fig:what-we-have}) and then deploy, test and use them, should be made accessible to larger audiences through leveraging the advantages of visual programming (Figure~\ref{fig:what-we-want}). This simple hook just accepts all transactions and logs that it is running. In the callback \textit{cbak} we define the emitted tx of a hook; this hook has no emitted tx and therefore no callbacks. \textit{"\_g(1,1)"}: every hook must import the guard function and use it at least once to guarantee that the execution will eventually terminate. \textit{"TRACESTR("Accept.c: Called.")"} and \textit{"accept(SBUF("Accepted!"),1)"} can be seen in \textit{log} and show the hook is working. Figure~\ref{fig:goal} shows why we advocate for VPLs: although showing the most basic \textit{hook}, even seasoned professionals could not understand it without explanations, if they were unfamiliar with \textit{hooks}.

In this paper we propose, implement and discuss \system, a solution for teaching non-technical audiences how to easily program, test and deploy some of the most challenging smart contracts like those written in C. To achieve this we use Visual Programming, a well-proven teaching methodology, and take the XRP Ledger smart contracts as a concrete case. As Google Blockly offers native support for the translation of the visual blocks to "friendlier" languages like JavaScript, Python, PHP, Lua, Dart, and the possibility to add other languages, other blockchains can be accommodated on this design too. The aforementioned languages are often considered to be \textit{higher-level} compared to C, with more abstraction and built-in functionality that makes them easier to learn and use for beginners.

The paper is organised as follows: Section~\ref{sec:bk} introduces the technologies involved. The~\system~solution is presented in Section~\ref{sec:design}, while the current state of the art is discussed in Section~\ref{sec:relwork}. Finally, we draw our conclusions and present the avenues for future work in Section~\ref{sec:conclusion}.

%% file: background.tex
% !TEX root = main.tex

\section{Background}
\label{sec:bk}

Here we are describing the most important technologies involved in~\system, like the \textit{XRP Ledger}, the \textit{Hooks} smart contracts on XRPL, and the \textit{Blockly} library for visual programming offered by Google.

\textbf{XRPL} is characterised as an open-source, permissionless, and decentralized blockchain which is appreciated for its transaction (tx) throughput (1500 tx/s)~\cite{ikeda2022first}, speed (transactions settle in 3-5s)~\cite{trestioreanu2023xrp}, low fees, and low energy consumption, all thanks to the consensus protocol involved: the ledger building process consists of a Byzantine Fault Tolerant \textit{"Consensus"}~\cite{security_analysis_xrpl} and a \textit{"Validation"} stage, where a majority of the participating nodes have to agree on the next version of the ledger. This is not a computationally-intensive process and it is designed to provide the above mentioned advantages. XRPL is focused on cross-border payments and has support for NFTs and for smart contracts which on XRPL are called \textit{Hooks}~\cite{hooks-xrpl}. 

\begin{figure}[ht]
\begin{center}
    \includegraphics[width=0.485\textwidth]{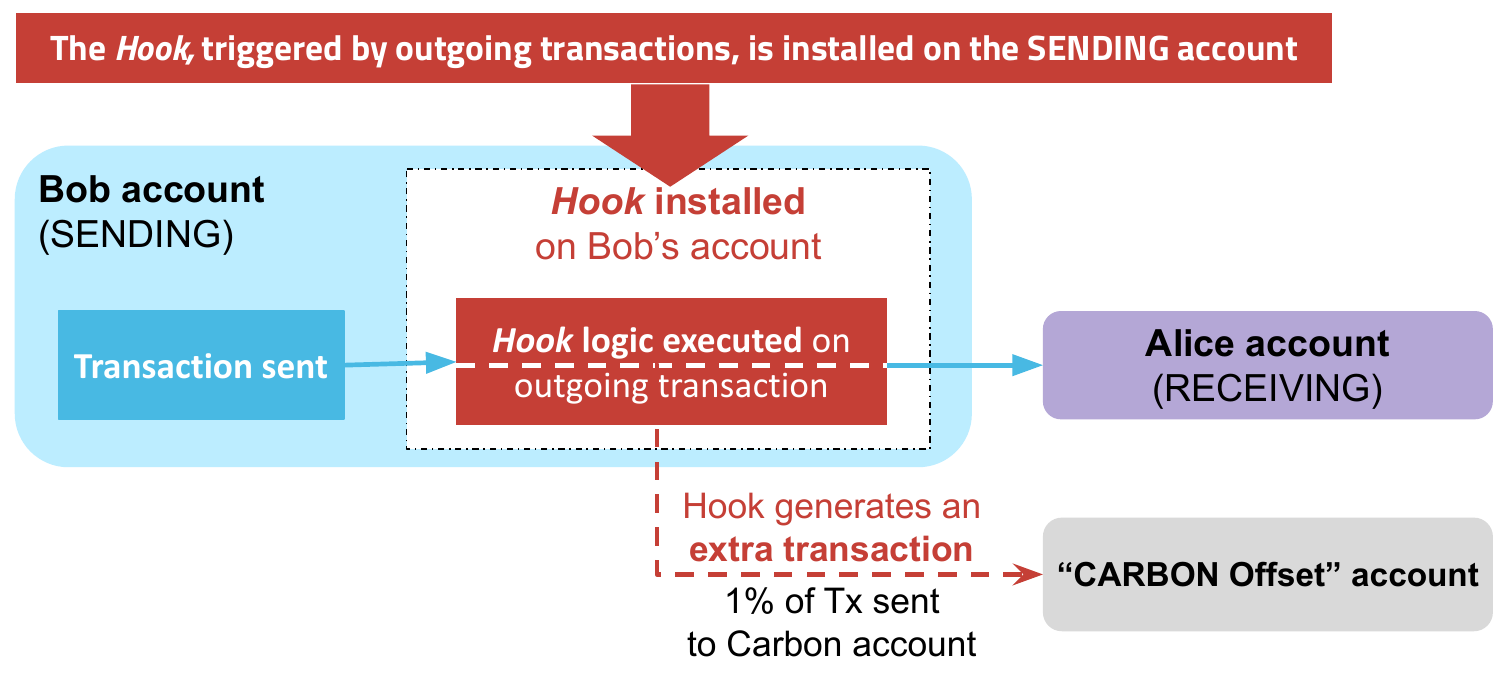}
    \caption{Example Hook on the XRP Ledger: "Carbon offset".}
    \label{fig:carbon-hook}
\end{center}
\end{figure}

The \textbf{Hooks} have been developed specially for the XRPL. They are small, efficient pieces of code compiled to web assembly (wasm) modules. Hooks can be written in any language (compilable to wasm) then they are uploaded to XRPL~\cite{hooks-testnet}. They are deployed and work on Layer 1, meaning directly on the XRPL, and their function is to modify the behavior and the flow of the transactions. The logic they deploy can be executed before or after the transactions. Hooks are deliberately made not to be Turing-Complete, which is undesirable at layer 1 because this would make it impossible to determine when the program would end. And without predictable maximum execution times, the XRPL might never advance to the next ledger. Hooks can store simple data objects like for example lists: \textit{"for all incoming payments, check if the sending account is in a black\textbf{list} (e.g. kept by another hook), and if yes: reject it”.} Other possible hook examples are: \textit{“deny transfers less than 20 XRP”}, or \textit{“for outgoing transfers, send xy\% to a predefined account”}. Typically, Hooks are written in C. While C is very efficient, it is not easy to learn and use for beginners. Even more, because Hooks are deliberately not Turing-Complete, they use a modified C language which makes it even more complicated for non-expert audience. An example of an XRPL Hook, called \textit{"Carbon Offset"}~\cite{hook-examples}, is illustrated in Figure~\ref{fig:carbon-hook}: this hook is installed on Bob's \textit{(sender)} account and is triggered by outgoing transactions from Bob's account: when Bob sends some funds to Alice, the Hook will trigger a new transaction that will send 1\% of the outgoing amount to a \textit{"Carbon Offset"} account. It is possible to install such hooks on other sending user accounts too, such that over time the "Carbon Offset" account would raise an amount that can be used to mitigate the effects of carbon emissions on \textit{Global Warming} through for example sponsoring reforestation.

\textbf{VPLs} are programming languages that use visual elements and graphical representations to create programs. Instead of writing code using text based syntax, users interact with graphical elements such as icons, symbols, and visual elements to build applications. Some key characteristics are: visual representation, drag-and-drop functionality, immediate feedback, visual debugging, rapid prototyping. They can be educational tools as well as be used to develop domain-specific applications. Examples include Scratch, Blockly, App Inventor, LabVIEW, or Max/MSP. Scratch~\cite{scratch} is a visual programming language developed at the MIT Media Lab. Scratch is used by thousands of kids worldwide to learn how to program and allows them to share their creations with one another over the Internet. The approach, its advantages and what made it so successful is thoroughly explained by the authors in~\cite{scratch-paper}. Nevertheless, Scratch is not easily customizable, and it only translates to Java Script. Another VPL example is Droplet~\cite{droplet} but it is not mature enough, nor widely used. Taleblazer~\cite{taleblazer} is a platform for developing Augmented Reality games using visual programming but it is specialized on gaming. As such, we found Google Blockly to be more suitable for our purpose:

\textbf{Blockly}~\cite{blockly} is an open-source framework from Google featuring visual block-based icons and a drag-drop programming environment. Non-expert users can leverage the  visual programming language approach provided by Blockly to build applications for education, gaming, robotics, or IoT. We chose Blocky because it can transform visual programs into many different textual codes, e.g., JavaScript, Php, Python, Dart, and Lua, and because it is much more versatile and customizable.

%% file: design.tex
% !TEX root = main.tex

\section{Methodology and result}
\label{sec:design}

\begin{figure}[ht]
\begin{center}
    \includegraphics[width=0.485\textwidth]{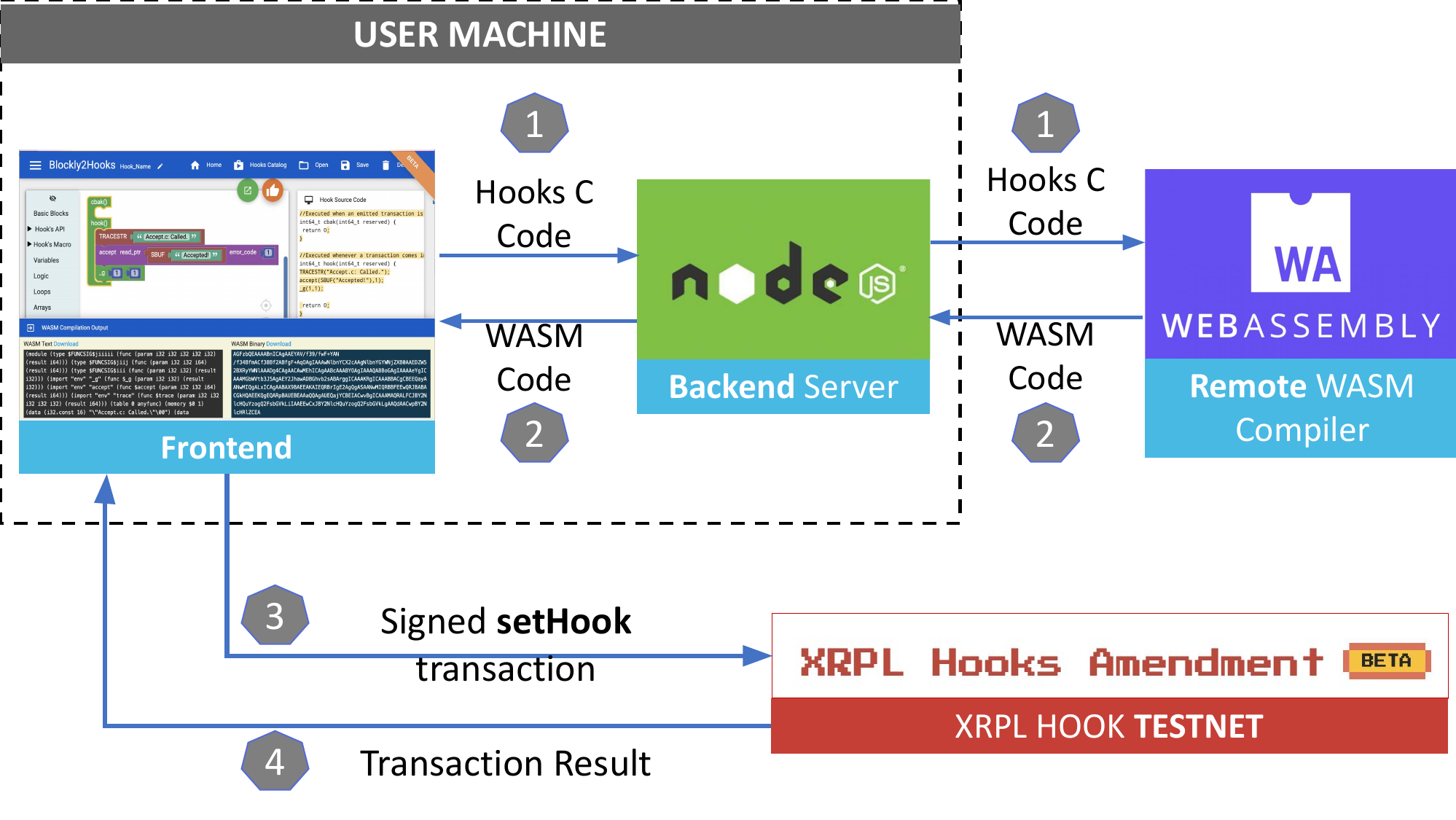}
    \caption{\system~architecture.}
    \label{fig:system}
\end{center}
\end{figure}

This section describes \system, the  proposed solution for bringing smart contracts closer to large, non-expert audiences. As stated, the goal is to enable everyone interested, to develop and deploy smart contracts even in challenging cases such as when the contracts are written in complex programming languages like C - as currently encountered for example on the XRP Ledger. We do this through leveraging the \textit{Visual Programming Language} (VPL) approach, more specifically by using the visual programming libraries offered by Google that are called \textit{Blockly}. To achieve this, we first study, map and implement the functions that we could identify in the XRPL Hooks smart contracts as visual \textit{blocks}, using Blockly. This enables the use of the “Drag \& Drop” visual programming approach for smart contract development. Next, we implement the compiling of the generated code to Web Assembly (WASM). Finally, through the push of a button after compilation, we enable users to easily sign and deploy to the blockchain, i.e. the XRP Ledger, the smart contract (hook) that they developed using our Blockly-based VPL. The proposed system architecture is represented in Figure~\ref{fig:system} and it comprises four major modules, which are described below:

%\begin{itemize}
%    \item The \textit{Frontend}
%    \item The \textit{Remote Server} for code compilation to web assembly
%    \item The \textit{Backend}
%    \item The \textit{XRPL Hooks Testnet}
%\end{itemize}

\begin{figure}[ht]
\begin{center}
    \includegraphics[width=0.485\textwidth]{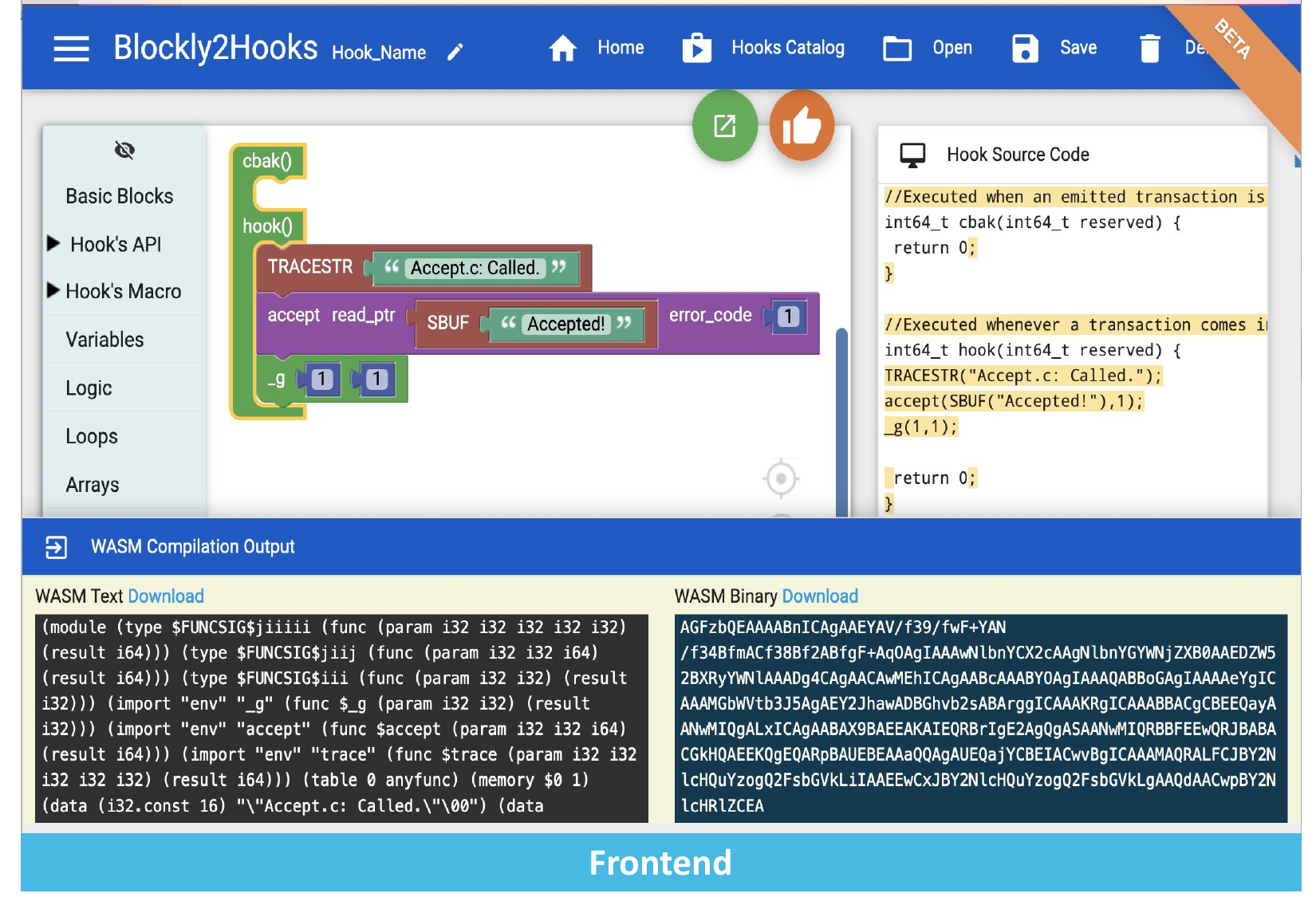}
    \caption{\system~Frontend.}
    \label{fig:frontend}
\end{center}
\vspace{-3mm}
\end{figure}

\textbf{M1. Frontend.} The user builds smart contracts through interacting with a web interface in their browser. The Frontend features a visual programming environment where users without advanced programming skills can \textit{drag and drop} blocks and fill in basic data (e.g. amounts) needed to build their desired \textit{Hook Smart Contract}. The result is translated to Hooks' C code and can be seen on the same screen on the Frontend, as illustrated in Figure~\ref{fig:frontend}. 

\textbf{M2. The remote wasm compiler server} receives the hooks that have been visually built by the user in the form of C code, and then compiles them to wasm, returning either the same hooks but in the wasm format, or the error(s) encountered during compilation. We chose to externalise the wasm compiler to keep the user experience light and smooth, because the wasm compiler involves the management of complex libraries required to compile the hooks from the special XRPL's C language format to the wasm format. As such, the user needs to fetch and install only the light code associated to the \textit{backend} and the \textit{frontend}. In production, there may be questions about trusting this server and its produced WASM code. In the proof-of-concept we used a remote compiler to keep the focus on our main question: using VLP to develop smart contracts. However, this compiler should be verified and trusted in production~\cite{security-wasm}. In future work we will tackle the potential security issues of smart contract compilation and deployment.

\textbf{M3. The XRPL Hooks Testnet} is a parallel XRPL network (not the production one) where users can learn and experiment with deploying hooks without risking to lose real money. The \textit{Testnet} also offers its own \textit{Faucet}, where users can get a \textit{Testnet} account and "fake" money (XRP).

\textbf{M4. The backend} is an interface between the Frontend and the Remote Wasm Compiler. After the \textit{Hook Smart Contract} is built, the user compiles the translated C code to \textit{wasm}: through the push of a button on the Frontend, the C code is  sent to the \textit{Remote Wasm Compiler Server}, to be compiled to \textit{wasm}. This process is managed by the \textit{Backend}, which takes the C code from the Frontend and forwards it to the \textit{Remote wasm compiler}. In turn, the \textit{Remote Wasm Compiler Server} will return to the Backend the hook as a wasm file or the compilation error(s). The Backend will forward the received hook, or the error(s), to the Frontend which will display them on-screen to the user.

For security reasons, the rest of the process is handled on the Frontend: after receiving the hook as wasm code, the user signs the transaction and sends it to the Testnet for deployment. This must happen on the Frontend because the signing process involves private credentials which in a real-life scenario should preferably remain at all times on the user's machine. After deploying the signed transaction, the user will get back to the Frontend the result of the transaction: either \textit{success}, or \textit{failure}.

\begin{figure}[ht]
\begin{center}
    \includegraphics[width=0.4\textwidth]{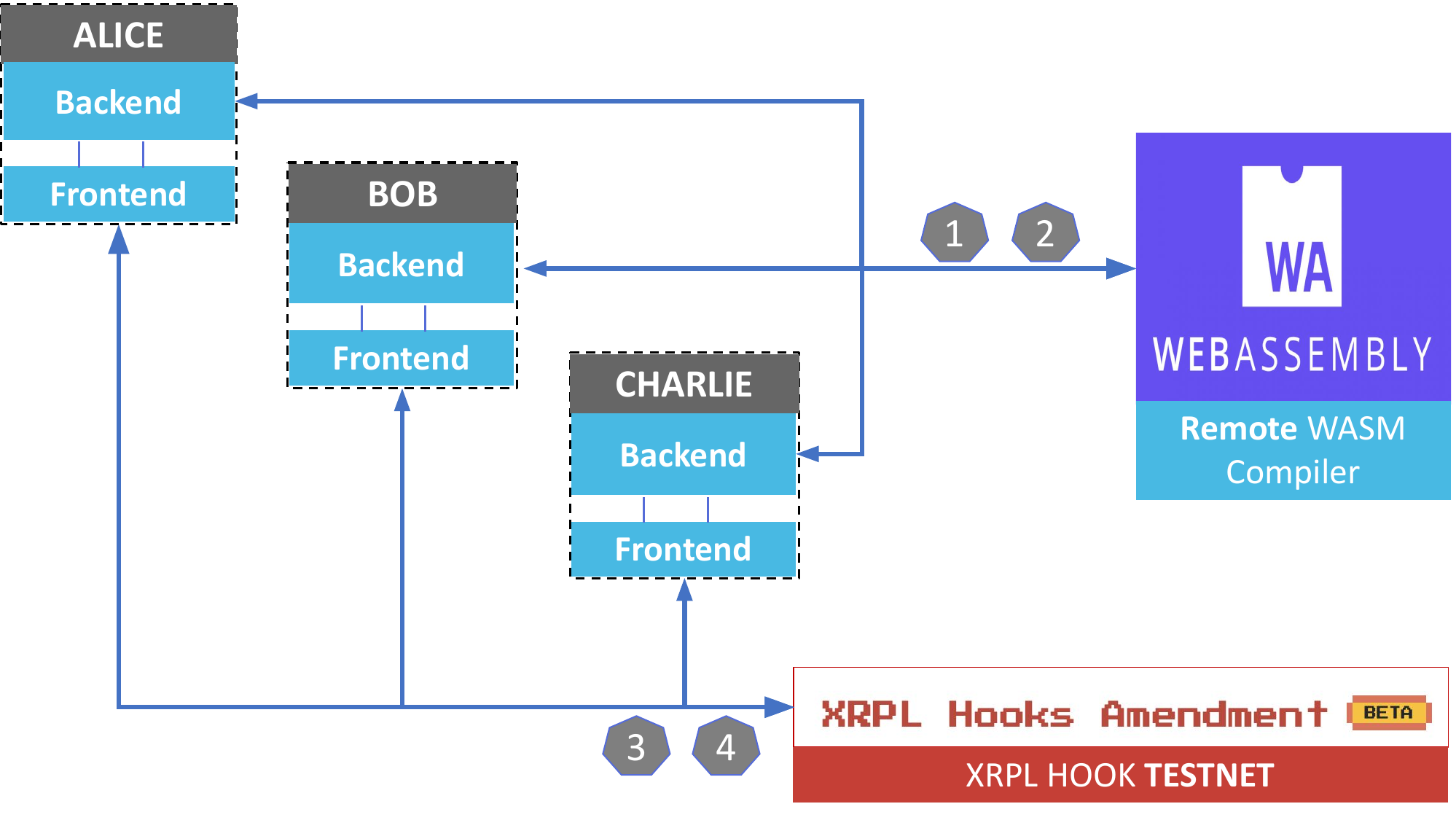}
    \caption{\system~multi-user architecture.}
    \label{fig:B2H-multiuser}
\end{center}
\end{figure}

The multi-user architecture is presented in Figure~\ref{fig:B2H-multiuser}. Users having installed on their local machines the free and open-source code consisting of the two \textit{Frontend} and \textit{Backend} \system~modules bundled together, will directly connect to the \textit{remote wasm compiler} and to the \textit{XRPL Hooks Testnet}, and will be instantly able to design and deploy Hooks. 
%Blockly2Hooks can also be easily connected to the XRPL Mainnet (the XRPL production network where real money are involved), to deploy the hooks to the Mainnet after testing them on the Hooks Testnet.
The platform is released open-source on GitHub\footnote{https://github.com/wshbair/blockly2hooks}. We have integrated some example Hooks for making it easier to get started and we are looking forward to adding more examples.

%% file: related_work.tex
\section{Related Work}
\label{sec:relwork}

Marks~\cite{marks} identifies different levels of abstraction for smart contract editors, from syntactic graphical editors such as those based on Blocky, through flow-based editors alike Unreal Engine \textit{Blueprint}, %~\footnote{https://docs.unrealengine.com/5.0/en-US/blueprints-visual-scripting-in-unreal-engine/, valid in February 2023}, 
and to \textit{forms} to be filled-in for the simplest cases. It argues that because the necessary level of abstraction is dependent on the intended audience and use-case, until the users are there, the abstraction level can only be guessed. The development of real-life contracts as smart contracts on Ethereum was studied in~\cite{detailed-sc-eth}, where challenges like the complexity of contract clauses and user privacy are exposed. The usage of visual domain-specific languages for smart contract creation on Solidity is studied in~\cite{DasContract,towards-skotnika-DasContract2} and a concrete solution named \textit{DasContract} is proposed. Actually, most of the previous work focused on visual editors for Solidity on Ethereum, possibly because this is the most popular smart contract platform: for example Latte~\cite{latte} provides feedback regarding the Gas cost incurred by the smart contract being built, while~\cite{ETH-Visual-Auto} employs a machine learning approach to build a visual programming environment. A solution based on YAWL~\cite{yawl} is proposed by \cite{industry-construction-users-themselves} for Solidity, while working on a construction industry use case. For an industry client, \cite{smart-contracts-for-purchases} designed and implemented a VPL-based environment for a \textit{legal purchase agreements} use case. The authors of~\cite{reuse-SC-visual} identify a lack of smart contract descriptors~\cite{reuse-SC-visual}, investigate the reusing of smart contracts, propose, then implement a design for smart contract descriptors, a descriptors registry, and a visual editor based on Google Blockly for creating composite smart contracts. The author of~\cite{purnell2022towards} identifies several requirements for widespread adoption of smart contracts in business and industry: ease of use, understandability, ease of testing, secure and error-free, scalable and affordable~\cite{purnell2022towards}. Next, it investigates declarative languages on a \textit{"Will and Testament"} use case.
\textit{FlowContract}\footnote{https://flowcontracts.com/docs, valid in February 2023} aims to be a flow-based editor for Solidity, which however continues to expose lots of low-level programming elements to the user.

\textit{Blocks}\footnote{https://blocks-editor.github.io/blocks/, valid in February 2023}~\cite{blocks-article} is an online editor for the \textit{Internet Computer}\footnote{https://dfinity.org/, valid in February 2023}. It is a flow-oriented design retaining many syntactic elements: most of the blocks, fields and variable names come from their text-programming counterparts. A design for Hyperledger Fabric is proposed in \cite{blockly-hyperledger}, however only limited work seems to be open-source. XRPL Labs proposes an online, browser-based \textit{Builder}~\cite{hooks-builder} for \textit{Hooks} which aggregates documentation, examples, and Testnet deployment. Smart contracts are built in C though, which ultimately makes it a programmer tool. Same as \textit{Builder}, \system~could eliminate the need to install code on user machines by placing online the functionalities still present on users' machines, such that users do everything from their browser by accessing a remotely-served web interface. To the best of our knowledge no other open-source project addresses the visual development of smart contracts natively developed with advanced languages such as the modified C for XRPL Hooks, and the environment proposed is general enough to accommodate any smart contract.

%% file: conclusions.tex
% !TEX root = main.tex

\section{Conclusions and Future Work}
\label{sec:conclusion}

For non-experts, developing smart contracts is currently a kind of mystery, with the field "reserved" to a narrow segment of seasoned professionals. In this work we investigated how interested audience with low programming skills can learn to develop smart contracts even like those written in advanced languages such as C. We propose a classic, proven methodology for teaching programming which is shown to achieve good results: the Visual block-based icons and a Drag\&Drop programming environment which is successfully used in schools to teach programming. As a concrete case we used the XRPL Hooks, and the visual programming environment we propose is built on Blockly, a comprehensive but in the same time flexible, friendly and future-proof framework which offers possibilities for extending and improving the project. 

Visual Programming is most likely not a silver bullet for smart contract development: for the foreseeable future, novel cases and complex smart contracts will most probably continue to be developed by highly skilled professionals. However we appreciate that visual programming could bring value to "industry" because it can help minimise the development time and costs for casual, repetitive use cases. Because the stakeholders build the smart contract themselves, the development time could be decreased also by eliminating the possible iterations required when the programmer misunderstands the exact need of the client. This workflow is already proven to work in practice: for example, misunderstandings between the game designer and the programmers happen in the video game industry, and besides lowering the development costs by freeing programmer time, it is a reason why flow-based visual editors have been introduced successfully in this industry. In case of a new high-usage use-case, the programmer(s) would just build the required new macro block(s). 

\system~was developed and tested on the Hooks Testnet and the results are promising to make learning smart contract development smoother. The demo is available online\footnote{https://bit.ly/blockly2hooks} and received good feedback from the community so far.

\textit{Future work.} At the time of writing, it appears that \textit{Hooks} is undergoing a security assessment; our platform will be connected to the Mainnet (the production network where real money are spent) as soon as \textit{Hooks} will be enabled there. Hence, users will be able to test their hooks on Testnet before deploying them on Mainnet. %Reverting the process to enable non-technical users to "read" smart contracts written as a computer program is also on our roadmap. 
As said before, tackling the potential security issues of smart contract compilation and deployment will be a priority. Additionally, we plan focus-testing the platform, to gather more feedback from the community to help improve \system~which is currently rather a syntactic graphical editor oriented towards learning smart contract development. Nevertheless, the platform is flexible enough to accommodate, in parallel, higher levels of abstraction: e.g., for the next version we consider the implementation of templates, or macro-blocks, for some of the most sought after use cases. This will enrich the design to include a \textit{flow-like} programming experience through interconnecting macro blocks treated by the user as black boxes.